# The evolution of conditional moral assessment in indirect reciprocity

Tatsuya Sasaki[1*], Isamu Okada[2], Yutaka Nakai[3]

[1]Faculty of Mathematics, University of Vienna, Oskar-Morgenstern-Platz 1, Vienna, Austria

[2]Department of Business Administration, Soka University, 1-236 Tangi, Hachioji-city, Tokyo, Japan

[3]Faculty of Systems Engineering and Science, Shibaura Institute of Technology, Fukasaku 307, Minuma-ku, Saitama-city, Saitama, Japan

[*]Correspondence to: tatsuya.sasaki@univie.ac.at

## Abstract

Indirect reciprocity is a major mechanism in the maintenance of cooperation among unrelated individuals. Indirect reciprocity leads to conditional cooperation according to social norms that discriminate the good (those who deserve to be rewarded with help) and the bad (those who should be punished by refusal of help). Despite intensive research, however, there is no definitive consensus on what social norms best promote cooperation through indirect reciprocity, and it remains unclear even how those who refuse to help the bad should be assessed. Here, we propose a new simple norm called "Staying" that prescribes abstaining from assessment. Under the Staying norm, the image of the person who makes the decision to give help stays the same as in the last assessment if the person on the receiving end has a bad image. In this case, the choice about whether or not to give help to the potential receiver does not affect the image of the potential giver. We analyze the Staying norm in terms of evolutionary game theory and demonstrate that Staying is most effective in establishing cooperation compared to the prevailing social norms, which rely on constant monitoring and unconditional assessment. The application of Staying suggests that the strict application of moral judgment is limited.

# Introduction

How to get unrelated people to mutually cooperate is a fundamental issue in today's highly mobile society. The last decades have seen researchers exploring indirect reciprocity, which is a major mechanism in the maintenance of cooperation between non-relatives[1-10]. The emergence of cooperation by indirect reciprocity can be summarized as, "I will help you if you have helped someone"[10]. Because helping is costly, however, self-interested recipients of help tend to freeload off others without further reciprocation, and unconditional cooperation is unlikely to evolve unless a specific supportive mechanism is provided[11]. Conditional cooperation, a main paradigm for exploring cooperation[12], suggests that cooperation should be channeled to those who deserve help by using social network assessment systems, such as reputation or gossip media[13-18].

How should one assess others' past behaviors? The simplest social norm, called Scoring, assesses those who give and refuse to give help as good and bad, respectively[19,20]. This norm depends only on individuals' previous actions. Since the seminal study of Nowak and Sigmund[10], Scoring has been investigated in terms of evolutionary game theory, primarily using a donor–recipient giving game.

Unconditionally applying the Scoring norm raises a key question: Is it morally or socially acceptable to refuse to help someone with a bad image? This point is Scoring's Achilles' heel in the typical good-or-bad binary system[10]. By definition, when a discriminator refuses to help a potential opponent with a bad image, the discriminator's image decisions become clouded. Thus, in the Scoring norm, a bad image is contagious. Although helping can redress the discriminator's image, a bad image may cause the discriminator to undergo rejection by other discriminators. Even slight involuntary errors can damage a discriminator's image and thus payoff. The Scoring norm, therefore, results in the mutual defection of all players[21,22]; this is referred as the "Scoring dilemma."

To address this dilemma, social norms have been developed that distinguish between justified and unjustified defection by accounting for the recipient's image[1]. In the case of a bad recipient being refused help, such refusal should not damage the donor's image (i.e., justified defection)[1,21,22]. Indeed, the top eight social norms, identified from 4,096 candidate strategies by systematic research (called the "leading eight")[23,24] share a common relevant feature—if a good donor refuses to help a bad recipient, the donor is assigned a good image (see Table S1).

Although the leading social norms are highly sophisticated, they are thus cognitively costly. Indeed, all of the leading eight rely on (i) the donor's last action and (ii) the recipient's last image (i.e., second-order social norms), and six of them also rely on (iii) the donor's last image (i.e., third-order social norms)[25]. Accounting for the image of both players would be rational in theory yet may overtax individuals in practice. Empirical studies on indirect reciprocity games have reported experimental results that many participants' assessments have appeared to rely only on the player's actions, not also on the images a player views[26]. We posit that there is a simpler method than that provided by the leading eight.

To tackle the Scoring dilemma, we formulate an opposite approach from prior paradigms by not applying higher-order social norms. We consider the effects of ignoring anxiety-producing stimuli from controversial interactions ("selective inattention"[27]). Assessing a donor who interacts with a recipient who has a bad image is a difficult task and is likely to lead an observer to feel regret about his or her assessment, regardless of whether the assessment is determined to be good or bad. This stressful situation may lead an observer to intuitively prefer inattention to interaction, that is, to abstain from making assessments. There is much supportive evidence of inattentive behavior by experiments[28,29] and field research[30-32]. Recent experimental data also suggest that a substantial fraction of subjects selectively consider the information of donor and recipient (that is, first- and second-order information) when making moralistic decisions[33]. Thus, we assume conditional observation on the opponent's image[34,35] (Fig. 1). The assessment system abstains from observation and thus also from assessment of whether or not refusal of help to the bad recipient is justified. Conditional assessment can prevent damage to the donor's image as it substantially generalizes the standard framework of indirect reciprocity to a meta-choice of {Assess, Preserve}. In the Assess case, the assessment is made according to the specific social norm, whereas in the Preserve case, the pre-existing image of the focal player is kept as is. The Preserve option is applicable for a broad range of assessment systems. We apply a conditional assessment for Scoring, leading to a new social norm—to perform an assessment as Scoring when a potential recipient has a good image; otherwise, to abstain from assessment. We call this new norm Staying.

# Results

We model our paradigm on the giving game in which the donor player has an opportunity to help the recipient player at a personal cost $c>0$; if the donor helps, then the recipient earns benefit $b>0$, with $b>c$. For simplicity, we assume that all discriminators share the same information about all personal images provided by this unique assessment system. We then consider both (i) implementation error[21,36], by which intentional help involuntarily fails with probability $e_1$, and (ii) assessment error[25], by which the assessment system mistakenly assesses a good donor as bad or a bad donor as good, with probability $e_2$.

We compare Staying with the four most prevailing social norms[10,37,38]: Scoring[19,20,39-41], Simple-standing[1,21,22], Stern-judging[3,37], and Shunning[5,42] (Table 1). Simple-standing and Stern-judging are the only two second-order social norms among the leading eight. When a donor refuses to help a bad recipient, Simple-standing and Stern-judging assess the donor as good, whereas Scoring and Shunning evaluate the donor as bad. Simple-standing, the most tolerant norm, assigns a good image to a person helping an individual evaluated as bad, and under Shunning, the most strict norm, a bad image is assigned to a donor who does not help a potential recipient evaluated as bad. In contrast with the four most prevailing social norms, no assessment is performed under Staying; the player's pre-existing image is simply preserved.

To study the evolutionary effects of these different social norms, we assume distinct strategies in very large populations: cooperators, defectors, and discriminators. Cooperators unconditionally give help; defectors unconditionally refuse to help; and discriminators, irrespective of the social norm, give and refuse help to good and bad recipients, respectively. In this study, we investigate the replicator dynamics describing the tendency whereby strategies that result in above-average earnings grow in frequency[39,43]. We note that the three homogeneous population states of cooperators, defectors, and discriminators are trivial equilibria of the replicator dynamics. We further assume that image updating is much faster than the time scale of game interactions so that we can study the replicator dynamics at a stationary state of the image system[43]. More detail is provided in Methods.

We first describe the Staying paradigm (Fig. 2a). Staying results in a defector–discriminator mixed equilibrium R with the fraction of discriminators

$$z_R = \frac{e_2 c}{\lambda(b-c)}, \tag{1}$$

where $\lambda = (1-e_1)(1-2e_2)$ describes social visibility, the probability that a donor's intention to cooperate is clearly recognized by an observer. There is no interior equilibrium that consists of all three strategies. Equilibrium R consists exclusively of defectors and discriminators and is a unique non-trivial equilibrium. Equilibrium R is unstable, and its $z_R$ value indicates the minimal frequency of discriminators required to invade then take over a population of defectors. For a sufficiently small assessment error $e_2$, equilibrium R appears in the state space, and depending on the initial conditions, the ending population consists exclusively of either defectors or discriminators. As the assessment error $e_2$ gets close to 0, equilibrium R approaches the state of 100% defectors, and the range of initial conditions leading the population to evolve to the state of 100% discriminators expands (see Methods for detailed analysis).

The other four rules examined are Scoring, Simple-standing, Stern-judging, and Shunning (Fig. 2b-e). In these cases, for a sufficiently small cost–benefit ratio $c/b$, the evolutionary dynamics result in an unstable mixed equilibrium R of defectors and discriminators, with the fraction of discriminators

$$z_R = \frac{c}{\lambda b} \geq \frac{c}{b}. \tag{2}$$

This value is common across Scoring, Simple-standing, Stern-judging, and Shunning, yet generically different from that of Staying in equation (1). The evolutionary dynamics with Scoring, Simple-standing, Stern-judging, and Shunning are all bi-stable, similarly to that with Staying (see Supplementary Information, Text S1 for detailed analysis). In Simple-standing, Stern-judging, and Shunning, similarly, there is no interior equilibrium, and R is a unique non-trivial equilibrium. In contrast to this, in Scoring there can be a continuum of equilibria in the interior state space, and R is an end of the continuum (Fig. 2b).

It follows from equations (1) and (2) that for $e_2 < 1 - c/b$, the range of initial conditions leading to selecting for discriminators between the two strategies is wider for Staying than for the other four norms. Note that in any case of Scoring, Simple-standing, Stern-judging, and Shunning, a fraction of discriminators less than $c/b$ is incapable of invading successfully. In striking contrast to this, for a sufficiently small assessment error $e_2$, Staying can thrive even if the initial fraction of discriminators is very low taking over the population of defectors, irrespective of the cost–benefit ratio.

**Discussion**

Most theory on the evolution of cooperation by indirect reciprocity is based on unconditional assessment. Evolutionary study on conditional assessment has started mainly by individual-based simulations, in which the corresponding assessment rule was named "us-TFT"[34,35]. In this paper, we fully analyze Staying, which is characterized by conditional assessment, and reveal that discriminators with Staying are more likely to invade the population of defectors than those with the four most prevailing social norms of indirect reciprocity[10,37,38]: Scoring[19,20,39-41], Simple-standing[1,21,22], Stern-judging[3,37], and Shunning[5,42] (Table 1). In mutual defection, within the population of defectors, either Simple-standing or Stern-judging leads defectors evaluated as bad to look good and then to exploit help from other discriminators. In contrast, under the Scoring or Shunning norm, discriminators are evaluated as bad as a result of interacting with defectors, leading to rejection by other discriminators; this is the main reason why the four social norms Scoring, Simple-standing, Stern-judging, and Shunning are unlikely to emerge. In contrast, Staying can leave the images of good discriminators and bad defectors intact; this enables discriminators to channel their cooperation and subvert the stalemate of mutual defection even with a small perturbation of the population state (which is on the order of assessment errors; see equation (2)).

The advantages of Staying are not limited to the emergence of effectiveness. A third-order social norm in the leading eight, called Strict-standing[23] (or L7 [43]), can provide the same image dynamics as Staying (see Table S1). This fact indicates that Strict-standing is as evolutionarily stable as Staying; however, Strict-standing and Staying are conceptually different. Staying preserves the donor's image as a result of abstention from observation; in contrast, Strict-standing reassigns the previous image during execution of the observation. The leading eight, which share the obligatory nature of unconditional observation, are less advantageous than Staying in terms of the load and error of cognitive process and information transfer. As such, conditional assessment, a property that characterizes Staying, can facilitate savings on the cost associated with running moral assessment systems. This would help mitigate the problem of second-order free riders who fail to contribute to costly assessments and only rely on assessment information provided or financed by others[44,45].

Our model can serve as a point of departure for investigating the effects of conditional assessment in various situations relevant to the evolution and development of human cooperation. Our model resembles a simplified top-down situation in which a unique media source delivers a story to infinitely large followers, who share the same assessment. The top-down situation considered has been extended in investigating situations in which assessment criteria and perception process vary by different individuals[46,47], populations are finite and structured[48-51], and individuals often interact by both direct and indirect reciprocity[52,53], as in small villages with gossip systems. Also, advances in information technology (IT) would promote the broad use of assessment systems with multivalued scores[19,54] or continuously varying scores[55], which otherwise usually require more cognitive load than do the binary assessment models considered. Thus, the results of future work that examines conditional assessment in indirect reciprocity by local gossip networks or sophisticated IT-aided rating, as such, will be fascinating.

Refusing to help the bad is difficult to morally assess. The situation causes a variety of controversial opinions and, viewed as a kind of punishment, is related to the argument about the need to consider the reputation of the punisher[56-63]. Previous study of indirect reciprocity answered strictly to this situation with a clear standard of justice; in contrast, Staying suspends the application of a scoring rule to this situation. Staying can be seen as a social norm applied loosely to some extent. As is known, the application of the law is difficult, and judicial discretion sometimes has to work. The controversy has been continuing between two principles. One recognizes that a law is sometimes forced to be applied loosely and tolerates judgment based on the judge's belief[64]. The other inhibits the judicial discretion and requires a judge to apply a law strictly[65]. We unveiled the excellence of Staying in forming social order. The findings suggest a limitation of the strict application of rules. In this sense, our study implies that the evolutionary study of indirect reciprocity can contribute to a further understanding of social norms and the law.

# Methods

**Indirect reciprocity in the giving game.** The main model is based on the standard framework for the evolution of indirect reciprocity[23-25]. Using this framework, discriminators are given a strategy by an assessment rule (called social norm) combined with an action rule. We base indirect reciprocity on the giving game, which is a two-player donation game in which one player acts as a donor and the other a recipient. The donor is given the opportunity to choose to help the recipient at a personal cost. The recipient can only receive help from the donor, if any is forthcoming. In other words, there is no option to reject help.

In this giving game, action rules prescribe to discriminators who are acting as potential donors how to respond to a potential recipient in a specific situation depending on the last image scores of both donor and recipient. We consider a simple model in which each individual is endowed with a binary image score of "good" or "bad." The action rule we apply to discriminators is to give help to a good recipient or to refuse to help a bad one, unless otherwise specified. After observing every interaction in the giving game, discriminators assign the donor's image by following the specific social norm, which is a function of (i) the donor's last action, (ii) the recipient's last image, and (iii) the donor's last image. When depending only on (i), the rule is called first order, when depending on both (i) and (ii), it is called second order, and when depending on (i), (ii), and (iii), it is called third order[23].

**Conditional assessment.** In the present study, we examine conditional assessment[34,35], which specifies a meta-choice of "Assess" or "Preserve." When choosing Assess, discriminators assign either a good or a bad image to a potential donor; when choosing Preserve, discriminators abstain from assessment. Based on this concept of conditional assessment, we introduce "Staying," which is a new social norm specifying that when a potential recipient is good, the donor's image should be assessed, as is done under Scoring, and when a potential recipient is bad, the donor's image should be preserved (that is, left unchanged).

To understand Staying, we also comparatively explore all first- and second-order social norms that take into account the recipient's last image and the donor's last action[10,37,38]. In particular, we focus on the four most prevailing social norms: Scoring[19,20,39-41], Simple-standing[1,21,22], Stern-judging[3,37], and Shunning[5,42]. Scoring is the best-known first-order social norm, and the other three are second-order social norms. Table 1 and Table S1 provide full details of these social norms. When assessing the use of Staying and the four norms, the consensus is that if a recipient has a good image, a good image should be assigned to those who helped and a bad image to those who did not help, the same as in the simplest case, Scoring.

**Observation, information, and errors.** In the model, we consider both public information and indirect observation. We assume indirect observation as players with the same social norm adopting and equally sharing the same image of a focal player, which has been provided by a representative observer. For simplicity, we also assume perfect information in which the probability that players know the image of a potential recipient is 100%. In addition, we consider both implementation error[21,36,38] and assessment error[25]. We denote by $e_1$ the probability that a player who intends to give help but fails to do so. On the other hand, for a player who intends to refuse to help, no implementation error occurs in the intentional refusal, and the player inevitably does so. The unilaterality of the implementation error, as such, can occur, in particular, when there is a lack of resources for helping, in which case it matters for an individual who intends to help (leading to the "phenotypic defector"[36]), yet does not for an individual who intends to refuse to help. We denote by $e_2$ the probability that an observer mistakenly assigns a bad image to a donor who should get a good one or assigns a good image to a donor who should get a bad one. The assessment error $e_2$ can be considered small enough ($<1/2$). The combination of the implementation and assessment errors is often considered in the theoretical study of the leading social norms for the evolution of cooperation by indirect reciprocity[23-25].

**Evolutionary dynamics.** To study the evolutionary dynamics of discriminators, we employ a continuous-entry model. An individual's social learning (or birth and death) sometimes happens, and when it does, this changes the strategy distribution in the population[39,58]. For analytic simplicity, we consider that in an individual's lifetime, that individual infinitely plays the one-round giving game with different opponents. We also consider an infinitely large population to examine replicator dynamics[43], which, in general, are described as $dx_S/dt = x_S(P_S - P)$, where $x_S$ denotes the relative frequency of strategy $S$; $P_S$ is the expected payoff for strategy $S$, given by the limit in the mean of the payoff per round for the strategy; and $P$ is the average payoff over the population, given by $\sum x_S P_S$. We note that each homogeneous state with $x_S = 1$ is a trivial equilibrium of the replicator dynamics. First, we examine three strategies: discriminators [S = Z], cooperators [X], and defectors [Y]. Cooperators unconditionally intend to help a potential recipient, and in contrast, defectors unconditionally intend not to help a potential recipient.

**Image dynamics.** To describe the dynamics of image scores, both good and bad, we use $g_S$ to denote the frequency of individuals with a good image among individuals adopting the same strategy $S$, and we use $g$ to denote the average fraction of individuals with good images over the population; thus, $g = xg_X + yg_Y + zg_Z$. In addition, we use $g_{S,I}$ to denote the probability that a good image is assigned to a potential donor who adopts strategy $S$ and also faces a potential recipient with an image score $I$ = good [G] or bad [B]. The population size is very large, so we assume that the composition of the population does not change between consecutive one-round giving games[39,43]. Thus, the frequencies of good players satisfy

$$g_X = g_{X,G}g + g_{X,B}(1-g),$$
$$g_Y = g_{Y,G}g + g_{Y,B}(1-g), \quad (3)$$
$$g_Z = g_{Z,G}g + g_{Z,B}(1-g).$$

**Staying norm.** We begin by analyzing the Staying norm, in which case equation (3) is described as

$$g_X = \varepsilon g + g_X(1-g),$$
$$g_Y = e_2 g + g_Y(1-g), \quad (4)$$
$$g_Z = \varepsilon g + g_Z(1-g),$$

where $\varepsilon = (1-e_1)(1-e_2) + e_1 e_2$. In equation (4), the first term $\varepsilon g$ in the sum for $g_X$ or $g_Z$ describes the probability that a cooperator or discriminator who faces a good recipient (probability $g$) is assigned a good image by giving help with no errors (probability $(1-e_1)(1-e_2)$) or by failing to give help with both errors (probability $e_1 e_2$). For $g_Y$, the term $e_2 g$ in the sum expresses the probability that a defector who faces a good recipient (probability $g$) is assigned a good image through assessment errors (probability $e_2$). The second term in the sum, $g_S(1-g)$, describes the probability that a donor who faces a bad recipient (probability $1-g$) is assigned a good image. In this case, according to the definition of Staying, the probability of finding a good discriminator should remain unchanged as $g_S$. Solving these equations leads to $g_Y = e_2$, $g_X = g_Z = \varepsilon$, and thus $g = e_2 y + \varepsilon(1-y)$.

Then, the expected payoffs are given by

$$P_X = (1-e_1)b(x+g_X z)-(1-e_1)c,$$
$$P_Y = (1-e_1)b(x+g_Y z), \quad (5)$$
$$P_Z = (1-e_1)b(x+g_Z z)-(1-e_1)cg.$$

This yields

$$P_Z - P_Y = (1-e_1)[b(g_Z - g_Y)z - cg]. \quad (6)$$

For $x=0$, that is, on edge YZ in Fig. 2a, we have

$$P_Z - P_Y = (1-e_1)[(\varepsilon - e_2)(b-c)z - e_2 c]. \quad (7)$$

This results in $z = z_R$ such that it satisfies $P_Z - P_Y = 0$, leading to

$$z_R = \frac{e_2 c}{(\varepsilon - e_2)(b-c)} = \frac{e_2 c}{(1-e_1)(1-2e_2)(b-c)}. \quad (8)$$

Considering that $0 < e_2 < 1/2$ and $b > c$, a boundary equilibrium R with $(x,y,z) = (0, 1-z_R, z_R)$ enters the edge YZ for the typical parameter settings ($b>c$, and $e_1$ and $e_2$ are sufficiently small). Along the edge, the replicator dynamics are described as $dz/dt = z(1-z)(P_Z - P_Y)$, and this yields $dz/dt < 0$ for $z < z_R$ or $dz/dt > 0$ for $z > z_R$. Equilibrium R is repelling along the edge and divides the edge into the basins of attraction for the homogeneous states of defectors ($y=1$) and discriminators ($z=1$).

Next, we turn to the payoff difference between cooperators and discriminators

$$P_Z - P_X = (1-e_1)[b(g_Z - g_X)z - c(g-1)]. \tag{9}$$

Considering $g_X = g_Z = \varepsilon$, it follows that

$$P_Z - P_X = (1-e_1)c(g-1) \geq 0, \tag{10}$$

where the average frequency of good $g$ is less than 1 when both errors are non-zero. This means that for the typical parameter settings, cooperators are dominated by discriminators and defectors, thus leading the population to converge to edge YZ ($x=0$). It follows that the homogeneous state of discriminators ($z=1$) is evolutionarily stable for sufficiently small errors (see Fig. 2a) and becomes globally stable when there is no assessment error, $e_2 = 0$.

## References

1. Sugden, R. *The Economics of Rights, Cooperation and Welfare* (Basil Blackwell, Oxford, 1986).

2. Alexander, R. D. *The Biology of Moral Systems* (Aldine de Gruyter, New York, 1987).

3. Kandori, M. Social norms and community enforcement. *Rev. Econ. Stud*. **59,** 63–80 (1992).

4. Wedekind, C. & Milinski, M. Cooperation through image scoring in humans. *Science* **288,** 850–852 (2000).

5. Panchanathan, K. & Boyd, R. Indirect reciprocity can stabilize cooperation without the second-order free rider problem. *Nature* **432,** 499–502 (2004).

6. Milinski, M., Semmann, D. & Krambeck, H. -J. Reputation helps solve the 'tragedy of the commons'. *Nature* **415,** 424–426 (2002).

7. Ule, A., Schram, A., Riedl, A. & Cason, T. N. Indirect punishment and generosity toward strangers. *Science* **326,** 1701–1704 (2009).

8. Yoeli, E., Hoffman, M., Rand, D. G. & Nowak, M. A. Powering up with indirect reciprocity in a large-scale field experiment. *Proc. Natl. Acad. Sci. USA* **110,** 10424–10429 (2013).

9. Watanabe, T. et al. Two distinct neural mechanisms underlying indirect reciprocity. *Proc. Natl. Acad. Sci. USA* **111,** 3990–3995 (2014).

10. Nowak, M. A. & Sigmund, K. Evolution of indirect reciprocity. *Nature* **437,** 1291–1298 (2005).

11. Rand, D. G. & Nowak, M. A. Human cooperation. *Trends Cogn. Sci.* **17,** 413–425 (2013).

12. Fischbacher, U., Gächter, S. & Fehr, E. Are people conditionally cooperative? Evidence from a public goods experiment. *Econ. Lett.* **71,** 397–404 (2001).

13. Diekmann, A., Jann, B., Przepiorka, W. & Wehrli, S. Reputation formation and the evolution of cooperation in anonymous online markets. *Am. Sociol. Rev.* **79,** 65–85 (2014).

14. Cuesta, J. A., Gracia-Lázaro, C., Ferrer, A., Moreno, Y. & Sánchez, A. Reputation drives cooperative behaviour and network formation in human groups. *Sci. Rep.* **5,** 7843 (2015).

15. Gallo, E. & Yan, C. The effects of reputational and social knowledge on cooperation. *Proc. Natl. Acad. Sci. USA*. **112,** 3647–3652 (2015).

16. Milinski, M. Reputation, a universal currency for human social interactions. *Phil. Trans. R. Soc. B* **371,** 20150100 (2016).

17. Antonioni, A., Cacault, M. P., Lalive, R. & Tomassini, M. Know thy neighbor: costly information can hurt cooperation in dynamic networks. *PLoS ONE* **9,** e110788 (2014).

18. Antonioni, A., Sánchez, A. & Tomassini, M. Cooperation survives and cheating pays in a dynamic network structure with unreliable reputation. *Sci. Rep.* **6,** 27160 (2016).

19. Nowak, M. A. & Sigmund, K. Evolution of indirect reciprocity by image scoring. *Nature* **393,** 573–577 (1998).

20. Nowak, M. A. & Sigmund, K. The dynamics of indirect reciprocity. *J. Theor. Biol.* **194,** 561-574 (1998).

21. Leimar, O. & Hammerstein, P. Evolution of cooperation through indirect reciprocity. *Proc. Biol. Sci.* **268,** 745–753 (2001).

22. Panchanathan, K. & Boyd, R. A tale of two defectors: the importance of standing for

evolution of indirect reciprocity. *J. Theor. Biol*. **224,** 115–126 (2003).

23. Ohtsuki, H. & Iwasa, Y. How should we define goodness?—reputation dynamics in indirect reciprocity. *J. Theor. Biol*. **231,** 107–120 (2004).

24. Ohtsuki, H. & Iwasa, Y. The leading eight: social norms that can maintain cooperation by indirect reciprocity. *J. Theor. Biol*. **239,** 435–444 (2006).

25. Brandt, H. & Sigmund, K. The logic of reprobation: assessment and action rules for indirect reciprocation. *J. Theor. Biol*. **231,** 475–486 (2004).

26. Milinski, M., Semmann, D., Bakker, T. C. M. & Krambeck, H. -J. Cooperation through indirect reciprocity: image scoring or standing strategy? *Proc. Biol. Sci*. **268,** 2495–2501 (2001).

27. Sullivan, H. S. *The Interpersonal Theory of Psychiatry* (W. W. Norton & Co, New York, 1953).

28. Martin, D. Rational inattention in games: experimental evidence. http://ssrn.com/abstract=2674224 (2015).

29. Cheremukhin, A., Popova, A. & Tutino, A. A theory of discrete choice with information costs. *J. Econ. Behav. Organ*. **113,** 34–50 (2015).

30. Lacetera, N., Pope, D. G. & Sydnor, J. R. Heuristic thinking and limited attention in the car market. *Am. Econ. Rev*. **102,** 2206–2236 (2012).

31. De los Santos, B., Hortaçsu, A. & Wildenbeest, M. R. Testing models of consumer search using data on web browsing and purchasing behavior. *Am. Econ. Rev*. **102,** 2955–2980 (2012).

32. DellaVigna, S. Psychology and economics: evidence from the field. *J. Econ. Lit*. **47,** 315–372 (2009).

33. Swakman, V., Molleman, L., Ule, A. & Egas, M. Reputation-based cooperation: empirical evidence for behavioral strategies. *Evol. Hum. Behav*. **37,** 230–235 (2016).

34. Nakai, Y. & Muto, M. Evolutionary simulation of peace with altruistic strategy for selected friends. *Journal of Socio-Information Studies* **9,** 59–71 (2005).

35. Nakai, Y. & Muto, M. Emergence and collapse of peace with friend selection strategies. *J. Artif. Soc. Soc. Simul*. **11,** 6 (2008).

36. Lotem, A., Fishman, M. A. & Stone, L. Evolution of cooperation between individuals. *Nature* **400,** 226–227 (1999).

37. Pacheco, J. M., Santos, F. C. & Chalub, F. A. C. Stern-judging: A simple, successful norm which promotes cooperation under indirect reciprocity. *PLoS Comput. Biol*. **2,** e178 (2006).

38. Ohtsuki, H. & Iwasa, Y. Global analyses of evolutionary dynamics and exhaustive search for social norms that maintain cooperation by reputation. *J. Theor. Biol*. **244,** 518–531 (2007).

39. Brandt, H. & Sigmund, K. Indirect reciprocity, image scoring, and moral hazard. *Proc. Natl. Acad. Sci. USA* **102,** 2666–2670 (2005).

40. Berger, U. Learning to cooperate via indirect reciprocity. *Games Econ. Behav*. **72,** 30–37 (2011).

41. Nax, H. H., Perc, M., Szolnoki, A. & Helbing, D. Stability of cooperation under image scoring in group interactions. *Sci. Rep*. **5,** 12145 (2015).

42. Takahashi, N. & Mashima, R. The importance of subjectivity in perceptual errors on the emergence of indirect reciprocity. *J. Theor. Biol*. **243,** 418–436 (2006).

43. Sigmund, K. *The Calculus of Selfishness* (Princeton University Press, Princeton, 2010).

44. Suzuki, S. & Kimura, H. Indirect reciprocity is sensitive to costs of information transfer. *Sci. Rep*. **3,** 1435 (2013).

45. Sasaki, T., Okada, I. & Nakai, Y. Indirect reciprocity can overcome free-rider problems on costly moral assessment. *Biol. Lett*. **12,** 20160341 (2016).

46. Uchida, S. & Sigmund, K. The competition of assessment rules for indirect reciprocity. *J. Theor. Biol*. **263,** 13–19 (2010).

47. Uchida, S. Effect of private information on indirect reciprocity. *Phys. Rev. E* **82,** 036111 (2010).

48. Santos, F. P., Santos, F. C. & Pacheco, J. M. Social norms of cooperation in small-scale societies. *PLoS Comput. Biol*. **12,** e1004709 (2016).

49. Wang, Z., Wang, L., Yin, Z-Y. & Xia, C-Y. Inferring reputation promotes the evolution of cooperation in spatial social dilemma games. *PLoS ONE* **7,** e40218 (2012).

50. Nakamura, M. & Masuda, N. Groupwise information sharing promotes ingroup favoritism in indirect reciprocity. *BMC Evol. Biol*. **12,** 213 (2012).

51. Suzuki, T. & Kobayashi, T. Effect of tolerance of reputation-making norms on corporation in social exchanges: an evolutionary simulation on social networks. *Sociological Theory and Methods* **26,** 31–50 (2011).

52. Roberts, G. Evolution of direct and indirect reciprocity. *Proc. Biol. Sci*. **275,** 173–179 (2008).

53. Molleman, L., van den Broek, E. & Egas, M. Personal experience and reputation interact in human decisions to help reciprocally. *Proc. Biol. Sci*. **280,** 20123044 (2013).

54. Berger, U. & Grüne, A. Evolutionary stability of indirect reciprocity by image scoring. http://epub.wu.ac.at/4087/ (2014).

55. McNamara, J. M. & Doodson, P. Reputation can enhance or suppress cooperation through positive feedback. *Nat. Commun*. **6,** 6134 (2015).

56. Rockenbach, B. & Milinski, M. The efficient interaction of indirect reciprocity and costly punishment. *Nature* **444,** 718–723 (2006).

57. Kurzban, R., DeScioli, P. & O'Brien, E. Audience effects on moralistic punishment. *Evol. Hum. Behav.* **28,** 75–84 (2007).

58. Ohtsuki, H., Iwasa, Y. & Nowak, M. A. Indirect reciprocity provides only a narrow margin of efficiency for costly punishment. *Nature* **457,** 79–82 (2009).

59. dos Santos, M., Rankin, D. J. & Wedekind, C. Human cooperation based on punishment reputation. *Evolution* **67,** 2446–2450 (2013).

60. Raihani, N. J. & Bshary, R. The reputation of punishers. *Trends Ecol. Evol*. **30,** 98–103 (2015).

61. Raihani, N. J. & Bshary, R. Third-party punishers are rewarded, but third-party helpers even more so. *Evolution* **69,** 993–1003 (2015).

62. Jordan, J. J., Hoffman, M., Bloom, P. & Rand, D. G. Third-party punishment as a costly signal of trustworthiness. *Nature* **530,** 473–476 (2016).

63. Przepiorka, W. & Liebe, U. Generosity is a sign of trustworthiness—the punishment of selfishness is not. *Evol. Hum. Behav*. **37,** 255–262 (2016).

64. Hart, H. L. A. *The Concept of Law* (2nd ed) pp.272ff (Oxford Univerity Press, 1994).

65. Dworkin, R. *Taking Rights Seriously*, pp.22ff (Harvard University Press, 1977).

Supplementary Information file includes:
Text S1: Evolutionary dynamics for second-order social norms
Table S1: Social norms: Staying, Scoring, Shunning, and the leading eight

**Acknowledgments:** T.S. acknowledges the Austrian Science Fund (FWF), P27018-G11. I.O. acknowledges JSPS KAKENHI Grant, 16H03120. Y.N. acknowledges JSPS KAKENHI Grant, 16H03698.

**Author Contributions:** Y.N. developed the concept; I.O. initiated the project; T.S., I.O., and Y.N. designed and analysed the model; T.S. wrote the paper.

**Competing financial interest:** The authors declare no competing financial interests.

# Tables

**Table 1. How social norms make moral assessments in giving games.**

| Conditions | Recipient's image | G | G | B | B |
|---|---|---|---|---|---|
| | Donor's action | C | D | C | D |
| Assessment: What does the donor image look like? | Staying | G | B | **P** | **P** |
| | Scoring | G | B | G | B |
| | Simple-standing | G | B | G | G |
| | Stern-judging | G | B | B | G |
| | Shunning | G | B | B | B |

"G" and "B" describe a good and bad image, respectively. "C" and "D" denote an action to help and to refuse to help, respectively. "P" means the image of a donor remains unchanged ("Preserve").

# Figures

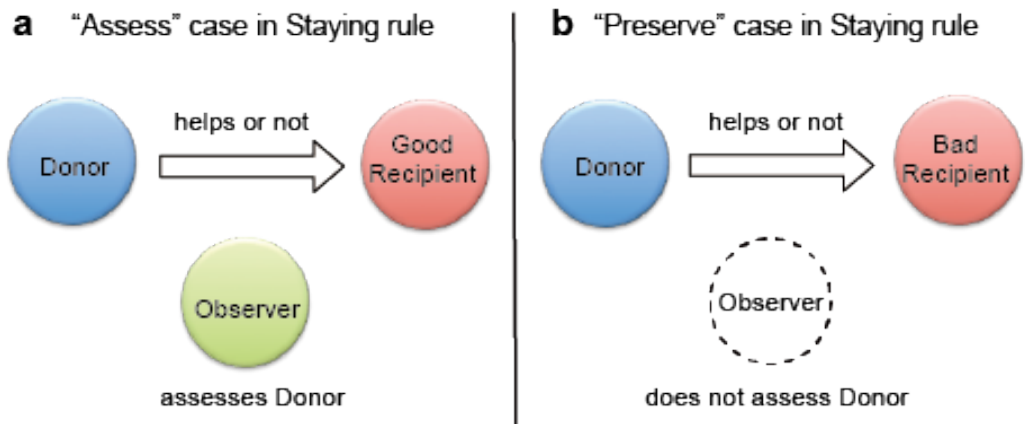

**Figure 1. Conditional assessment in giving games.** In the Staying rule, (a) the observer assesses the donor's image score, if the recipient has a good image; (b) otherwise, the observer does not assess the donor's image score, which stays the same.

**Table 2**

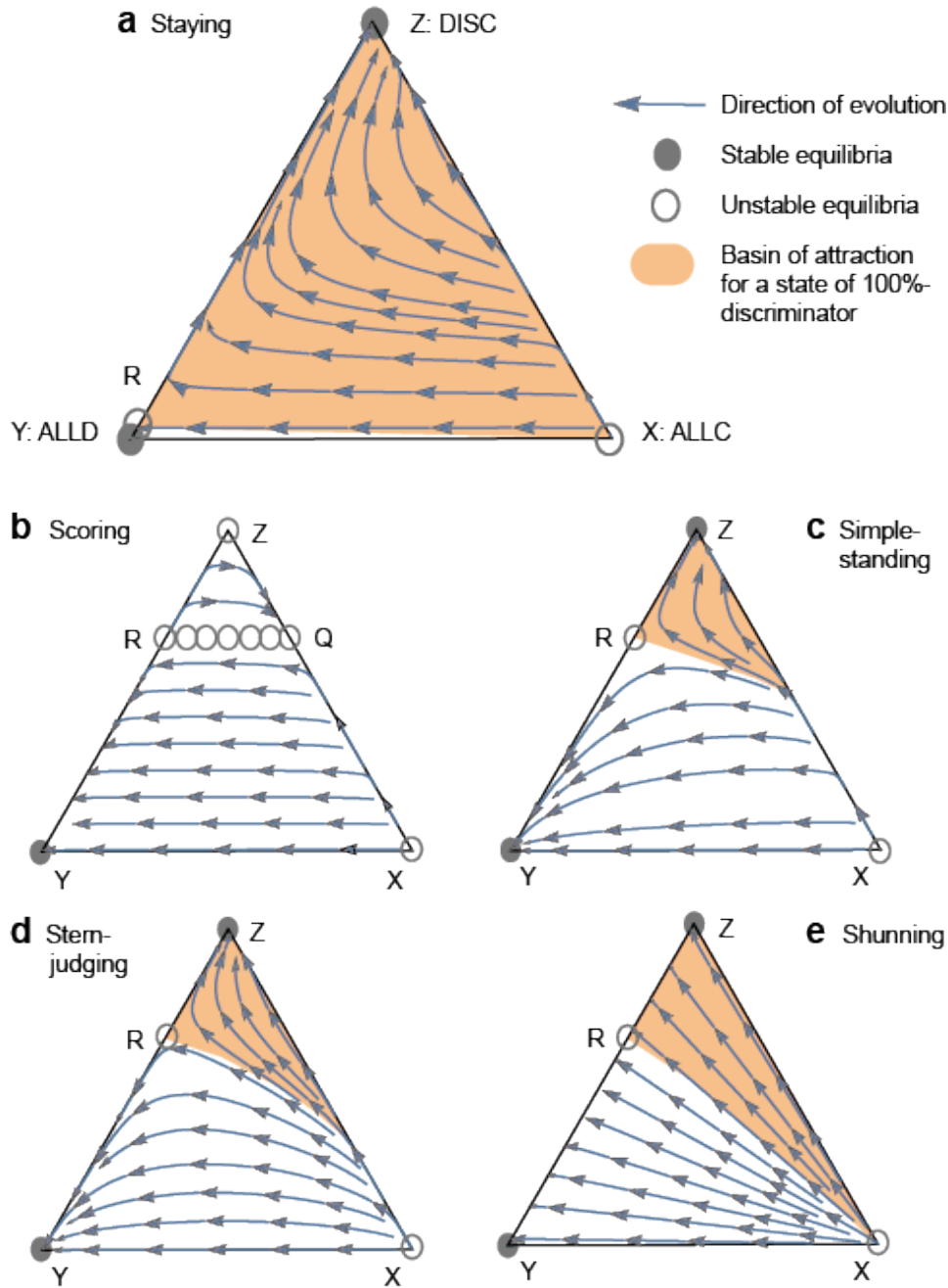

**Figure 2. Evolution of indirect reciprocity with different social norms.** The triangles describe a simplex of the state space $\{(x,y,z): x+y+z=1\}$, where $x,y,z \geq 0$ denote the frequencies of cooperators, defectors, and discriminators, respectively. Each node (X, Y, or Z: $x$, $y$, or $z=1$) of the triangle corresponds to the homogeneous state of each specific strategy. (a) Under Staying,

discriminators always are better off than cooperators. Thus, cooperators will vanish and the population will eventually converge to either node Z or Y. (b) Under Scoring, a continuum of equilibria connects boundary attractor Q and repeller R. The population drifts along the continuum and moves close to R, eventually attaining node Y. (c-e) Under Simple-standing, Stern-judging, or Shunning, the dynamics are qualitatively similar to those in (a). The basin of attraction for node Z is wider in (a) than in (b-e). Parameters: $c=1$, $b=1.5$, $e_1=e_2=0.01$. R corresponds approximately to $z_R=0.02$ in (a) or $z_R=0.66$ in (b-e).

# Supporting Information for

# The evolution of conditional moral assessment in indirect reciprocity

Tatsuya Sasaki, Isamu Okada, Yutaka Nakai

This PDF file includes Text S1 and Table S1.

## Text S1: Evolutionary dynamics for second-order social norms

We then focus on the 16 second-order social norms (except for one that prescribes unconditional defection given by BBBB (to unconditionally assign bad using Table 1). Compared to the case of the Staying norm, any of these 16 norms is less likely to invade a population of defectors. We note that since discriminators and defectors intend to refuse to help a bad recipient, $g_{Y,B} = g_{Z,B}$ holds. To analyze the replicator equations, we apply the expected payoffs, as in equation (5). For any of the 16 norms, equation (6) becomes

$$P_Z - P_Y = (1-e_1)[b(g_{Z,G} - g_{Y,G})z - c]g. \tag{S1}$$

We consider small but non-zero errors, which yield $0 < g < 1$. When $g_{Z,G} - g_{Y,G} \leq 0$, it follows that $P_Z - P_Y < 0$, and thus discriminators are dominated by defectors. When $g_{Z,G} - g_{Y,G} > 0$, this can result in, if any, a boundary equilibrium R with $z = z_R$, such that it satisfies

$$z_R = \frac{c}{(g_{Z,G} - g_{Y,G})b} \geq \frac{c}{b}. \tag{S2}$$

equation (S2) implies that for all cases of the 16 second-order social norms, the threshold frequency for discriminators to successfully invade a population of defectors is at least the cost-to-benefit ratio $c/b$.

**Scoring, Simple-standing, Stern-judging, and Shunning**

We compare the effectiveness of Staying with the results of the most prevailing social norms. We specifically check the global dynamics of Scoring, Simple-standing, Stern-judging, and Shunning. First, we note that by definition the conditional probability that a donor is assessed as good when a potential recipient is good (that is, the first term in the sum in equation (3)) is the same as it is in equation (4) of Staying. The difference is in the second term in the sum.

For **Scoring**, we obtain

$$\begin{aligned} g_X &= \varepsilon g + \varepsilon(1-g) = \varepsilon, \\ g_Y &= e_2 g + e_2(1-g) = e_2, \\ g_Z &= \varepsilon g + e_2(1-g). \end{aligned} \quad (S3)$$

Since Scoring is a first-order social norm that depends only on what a donor did, the degrees of goodness in cooperators and defectors, $g_X$ and $g_Y$, are independent of the recipient's degree of goodness $g$. A discriminator is assessed as good for a good recipient, with probability $\varepsilon$, or for a bad recipient, with the probability that he intentionally defects yet with assessment error $e_2$.

Next, for **Simple-standing**,

$$\begin{aligned} g_X &= \varepsilon g + (1-e_2)(1-g), \\ g_Y &= e_2 g + (1-e_2)(1-g), \\ g_Z &= \varepsilon g + (1-e_2)(1-g). \end{aligned} \quad (S4)$$

Simple-standing is the most tolerant norm, which is to assign a good image to a donor, irrespective of his/her actions to a bad recipient. Thus, the second term in the sum is the same as $(1-e_2)(1-g)$ over $g_X$, $g_Y$, and $g_Z$, in which case the donor is assessed as good only when no assessment error occurs.

Then, for **Stern-judging**,

$$\begin{aligned} g_X &= \varepsilon g + (1-\varepsilon)(1-g), \\ g_Y &= e_2 g + (1-e_2)(1-g), \\ g_Z &= \varepsilon g + (1-e_2)(1-g). \end{aligned} \quad (S5)$$

Stern-judging assigns a good image to those who refuse to help a bad recipient and a bad image to those who help a bad recipient. This leads to the second term in the sum for $g_X$. When a recipient is bad, unintentionally refusing help with no assessment error or intentionally giving help with assessment errors are both assessed as good. This conditional probability is $(1-e_1)e_2 + e_1(1-e_2) = 1-\varepsilon$.

Finally, for **Shunning**,

$$\begin{aligned} g_X &= \varepsilon g + e_2(1-g), \\ g_Y &= e_2 g + e_2(1-g), \\ g_Z &= \varepsilon g + e_2(1-g). \end{aligned} \quad (S6)$$

Shunning is the strictest case, in which a bad image is assigned to a donor irrespective of his/her actions toward a bad recipient. Thus, the second term in the sum is the same as $e_2(1-g)$ over $g_X$, $g_Y$, and $g_Z$, in which case the donor is assessed as good only when assessment errors occur.

Substituting equations (S3), (S4), (S5), or (S6) into equation (6), we obtain

$$P_Z - P_Y = (1-e_1)[(\varepsilon - e_2)bz - c]g. \quad (S7)$$

Considering $g > 0$, this results in a unique boundary equilibrium R with $z = z_R$, such that it is a repelling point and satisfies

$$z_R = \frac{c}{(\varepsilon - e_2)b} = \frac{c}{(1-e_1)(1-2e_2)b}. \quad (S8)$$

This leads to that when $(\varepsilon - e_2)b > c$, the unique equilibrium enters edge ZY ($x=0$). Comparing equations (8) and (S8), it is obvious that the basin of attraction for node Z ($z=1$) is wider under Staying than under the other four cases. Indeed, as the degrees of error, $e_1$ and $e_2$, move toward 0, the fraction necessary for discriminators to emerge, $z_R$ in equations (8) and (S8), converges to 0 and $c/b$, respectively. Thus, under Scoring, Simple-standing, Stern-judging, or Shunning, a sufficiently small cost-to-benefit ratio $c/b$ is required for rare mutants of discriminators to successfully invade a population of defectors. In striking contrast to this, under Staying, rare mutants of discriminators can invade as long as assessment errors $e_2$ are very small.

Next, we turn to the payoff difference between cooperators and discriminators in equation (9). For Scoring, substituting equation (S3) yields

$$P_Z - P_X = (1-e_1)[(e_2-\varepsilon)bz + c](1-g)$$
$$= -(P_Z - P_Y)\frac{1-g}{g}. \quad (S9)$$

Hence, there is a line consisting of fixed points with the same $z$-coordinate given by equation (S8). This line connects boundary fixed points Q on edge ZX ($y=0$) and R on edge ZY ($x=0$). We note that in contrast to R, Q is attracting along edge ZX. In particular, node Z is a saddle point, and node Y is a unique equilibrium that is asymptotically stable (see Fig. 2b). More details of these global dynamics can be explored by applying analogous arguments from refs. 5 and 39.

For Simple-standing or Shunning, substituting equation (S4) or (S6) yields the same results as in equation (10): $P_Z - P_X = (1-e_1)c(1-g) \geq 0$. For Stern-judging, substituting equation (S5) yields $P_Z - P_X = (1-e_1)[(\varepsilon-e_2)bz + c](1-g) \geq 0$. Thus, for the typical parameter settings, similarly to Staying, cooperators are dominated by discriminators and defectors, leading the population to converge to edge YZ ($x=0$). Thus, the global dynamics are qualitatively the same as those for Staying. However, the four social norms and the Staying norm differ quantitatively in the position of the repeller R, $z_R$ (see Fig. 2c-e).

|  | | Assessment: What does the donor image look like? | | | | | | | | Action: What should donor do? | | | |
|---|---|---|---|---|---|---|---|---|---|---|---|---|---|
| Conditions | Recipient's image | G | G | G | G | B | B | B | B | G | G | B | B |
| | Donor's image | G | B | G | B | G | B | G | B | G | B | G | B |
| | Donor's action | C | C | D | D | C | C | D | D | – | – | – | – |
| Scoring | | G | G | B | B | G | G | B | B | C | C | D | D |
| Shunning | | Same as in Scoring | | | | B | B | B | B | C | C | D | D |
| Leading eight | L1 (Standing) | | | | | G | G | G | B | C | C | D | C |
| | L2 | | | | | B | G | G | B | C | C | D | C |
| | L3 (Simple-standing) | | | | | G | G | G | G | C | C | D | D |
| | L4 | | | | | G | B | G | G | C | C | D | D |
| | L5 | | | | | B | G | G | G | C | C | D | D |
| | L6 (Stern-judging) | | | | | B | B | G | G | C | C | D | D |
| | L7 (Strict-standing) | | | | | G | B | G | B | C | C | D | D |
| | L8 | | | | | B | B | G | B | C | C | D | D |
| Staying | | | | | | P | P | P | P | C | C | D | D |

**Table S1. Social norms: Staying, Scoring, Shunning, and the leading eight.** "G" and "B" describe good image and bad image, respectively. "C" and "D" denote an action to help and to refuse to help, respectively. "P" means that the donor's image remains unchanged. With Scoring[19,20,39-41], whether to help or not determines the donor's image. When a potential recipient has a good image, the leading eight strategies[23,24] all have the same assessment as Scoring. Coding of the leading eight, L1 to L8, is the same as assigned in ref. 43. L1 is the original "Standing"[1,21,22], which is viewed as a third-order social norm. Only rules L1 and L2 have different action rules, which prescribe cooperation when both donor and recipient have bad

images. L1 (Standing) and L3 (Simple-standing) differ in that part of the action rule and also in the assessment of donors who refuse to help a bad recipient. Shunning[5,42] and Scoring do not belong to the leading eight, as Shunning could not achieve a sufficiently high degree of cooperation at the equilibrium state[38].